\documentstyle[aas2pp4,epsfig]{article}

\newcommand{\sfs}{$^1\mathrm{S}_0$~}
\newcommand{\sfp}{$^3\mathrm{P}_2$~}

\begin{document}

\title{Implications of Hyperon Pairing for Cooling of Neutron Stars}
\author{Christoph Schaab}
\affil{Institut f{\"u}r theoretische Physik,
  Ludwig-Maximilians Universit{\"a}t M{\"u}nchen, \\
  Theresienstr. 37, D-80333 M{\"u}nchen, Germany; \\ 
  schaab@gsm.sue.physik.uni-muenchen.de}
\author{Shmuel Balberg}
\affil{Racah Institute of Physics, The Hebrew University, Jerusalem 91904, 
        Israel;\\ shmblbrg@saba.fiz.huji.ac.il}
\and
\author{J{\"u}rgen Schaffner-Bielich}
\affil{Nuclear Science Division, Lawrence Berkeley National
  Laboratory, Berkeley, CA 94720;\\ schaffne@nta2.lbl.gov}

\begin{abstract}
The implications of hyperon pairing for the thermal evolution of
neutron stars containing hyperons are investigated. The outcome of
cooling simulations are compared for neutron star models composed only
of nucleons and leptons, models including hyperons, and models
including pairing of hyperons. We show that lambda and neutron pairing
suppresses all possible fast neutrino emission processes in not too
massive neutron stars. The inclusion of lambda pairing yields better
agreement with X-ray observations of pulsars.  Particularly, the
surface temperatures deduced from X-ray observations within the
hydrogen atmosphere model are more consistent with the thermal history
of neutron stars containing hyperons, if the critical temperature for
the onset of lambda and nucleon pairing is not too small.

\end{abstract}


\vspace{1cm}
\begin{center} \large
To be published in The Astrophysical Journal Letters
\end{center}

\section{Introduction}

Relativistic calculations of the composition of neutron star matter
lead to the conclusion that neutron stars are composed not only of
nucleons and leptons but also of hyperons and, possibly, of nucleon
isobars (see, e.g.,
\cite{Pandharipande71,Glendenning85a,Schaffner95a,Balberg97a,Huber97a}). It
was proposed by Prakash et~al. (1992) that the presence of hyperons
might lead to rapid cooling even if the proton fraction is too small
to allow for the nucleon direct Urca process and no exotic states
(like meson condensates and quark gluon plasma) exist
(see also \cite{Prakash94a}). This statement was confirmed by 
detailed cooling simulations (\cite{Haensel93,Schaab95a}).  The
role of hyperons in the cooling history of a neutron star is even
enhanced by the generally accepted neutron pairing in the interior of
the star (see, e.g., \cite{Takatsuka72,Amundsen85}). Whereas the
direct nucleon Urca is suppressed by neutron pairing (\cite{Lattimer94,Page95})
most of the hyperon processes are not. Haensel \& Gnedin (1994) and
Schaab et~al. (1996) found that unsuppressed rapid cooling due to
hyperon induced neutrino emission processes results in surface
temperatures that are too low compared with soft X-ray and extreme UV observations.

Recently Balberg \& Barnea (1998) calculated the \sfs gap energy of
$\Lambda$ hyperons in neutron star matter. The obtained gap energy is
similar to the \sfs gap energy of protons. As we will show here
hyperon pairing considerably changes the cooling behavior of a neutron
star containing hyperons (which we will call hyperon stars in the
following) and leads to better agreement with observations.

In this {\it Letter} we study the effect of hyperon pairing on the
thermal evolution of hyperon stars and compare the results with
``conventional'' neutron stars, which are composed of nucleons and
leptons only, and with observations. We begin with the underlying
physics by discussing the relevant neutrino emission processes and the
pairing of hyperons (Sect. \ref{sect:neutrino} and \ref{sect:gap}) and
describing in Sect. \ref{sect:eos} the relativistic equations of state
(EOS's) used. The results of the simulations are discussed in
Sect. \ref{sect:results}. In Sect. \ref{sect:conclusion} we compare
our results with observations and end with some conclusions.

\section{Neutrino Emission Processes} \label{sect:neutrino}

The early evolution of hot neutron stars is completely dominated by
the cooling via neutrino emission. Only after about $10^5$~yrs does
the photon radiation from the star's surface take over and dominate
the late evolution. This late cooling might be slowed down by various
heating processes (see \cite{Schaab97c} and references therein). The
neutrino emission processes can be divided into slow and enhanced
processes depending on whether one or two baryons participate. Due to
the rather different phase spaces associated with both kind of
processes the emission rates differ by several orders of magnitude.

\begin{table*}[tb] 
\caption{Possible direct Urca processes in hyperon stars
        \label{tab:neutrino}}
\smallskip
\begin{tabular}{cccccc}
& & \multicolumn{2}{c}{TM1-m1} & \multicolumn{2}{c}{RHF8} \\
Process & ${\mathcal R}_{B_1B_2}$-factor & $n_{\rm c}/n_0$ & $M_{\rm th}/M_\odot$ & $n_{\rm c}/n_0$ 
& $M_{\rm th}/M_\odot$ \\
\tableline
${\rm n}\rightarrow{\rm p}+{\rm e}^-+\bar\nu$   & 1      & 1.31 & 0.78  & 1.96  & 1.09  \\
$\Sigma^-\rightarrow\Lambda+{\rm e}^-+\bar\nu$  & 0.2055 & 1.89 & 1.30  & 2.99  & 1.35  \\
$\Lambda\rightarrow{\rm p}+{\rm e}^-+\bar\nu$   & 0.0394 & 1.90 & 1.31  & 3.11  & 1.38  \\
$\Sigma^-\rightarrow{\rm n}+{\rm e}^-+\bar\nu$  & 0.0125 & 2.37 & 1.45  & 2.44  & 1.22  \\
$\Xi^-\rightarrow\Lambda+{\rm e}^-+\bar\nu$     & 0.0175 & 2.65 & 1.52  & 6.80  & 1.65  \\
$\Sigma^-\rightarrow\Sigma^0+{\rm e}^-+\bar\nu$ & 0.6052 & 4.03 & 1.60  & 7.43  & $>M_{\rm max}$ \\
$\Xi^-\rightarrow\Sigma^0+{\rm e}^-+\bar\nu$    & 0.0282 & 4.03 & 1.60  &       &       \\
$\Xi^-\rightarrow\Xi^0+{\rm e}^-+\bar\nu$       & 0.2218 & 4.27 & 1.61  &       &       \\
$\Xi^0\rightarrow\Sigma^++{\rm e}^-+\bar\nu$    & 0.0564 & 4.30 & $>M_{\rm max}$&  & 
\end{tabular}
\tablecomments{${\mathcal R}_{B_1B_2}$ is the relative emissivity
$\epsilon_\nu(B_1 \rightarrow B_2+{\rm e}^-+\bar\nu)/\epsilon_\nu({\rm
n}\rightarrow{\rm p}+{\rm e}^-+\bar\nu)$, $n_{\rm c}$ the critical
density above which the process is allowed, and $M_{\rm th}$ the
corresponding threshold mass of a non-rotating neutron star (both
values depend on the underlying EOS, here: TM1-m1 and
RHF8).}
\end{table*}
In hyperon stars various enhanced neutrino emission processes can
occur, as listed in Tab. \ref{tab:neutrino}. The processes are only
noted in one direction. Because of $\beta$-equilibrium the inverse
reaction (with $\bar\nu$ replaced by $\nu$) occurs at the same
rate. The first process is the direct nucleon Urca process, the
remaining processes are comprised by the term ``direct hyperon
Urca''. The relative emissivities ${\mathcal R}_{B_1
B_2}=\epsilon_\nu(B_1 \rightarrow B_2+{\rm
e}^-+\bar\nu)/\epsilon_\nu({\rm n}\rightarrow{\rm p}+{\rm
e}^-+\bar\nu)$ of these processes are given with respect to the
emissivity of the direct nucleon Urca process (\cite{Prakash92}). 
Simultaneous conservation of energy and momentum
requires that the triangle inequality, $p_{B_1}^{\rm F} < p_{B_2}^{\rm
F} + p_{\rm e}^{\rm F}$, and the two inequalities obtained by cyclic
permutation are fulfilled for the Fermi momenta $p_i^{\rm F}$.
If the inequalities are not fulfilled
the process is extremely unlikely to occur and
the corresponding emissivity vanishes.
The availability of the various fast processes is thus dependent on
the partial concentrations of each baryon species, which in turn is
determined by the EOS.

If one of the participating baryons pairs in a superfluid state the
emissivity is suppressed by an approximately exponential factor:
${\mathcal R}_{\rm sf}\sim\exp(-cT_{\rm c}/T)$ for $T\ll T_{\rm c}$
where $T_{\rm c}$ is the critical temperature and
$c$ is a constant of the order unity.  Since the Fermi momenta of the
hyperons are sufficiently small they pair in a \sfs state (see
Sect. \ref{sect:gap}), as is the case for protons at slightly 
lower densities. Due to their high Fermi momenta, neutrons are
expected to pair in an anisotropic \sfp state. We refer to Levenfish
\& Yakovlev (1994) for fitting formulas of ${\mathcal R}_{\rm sf}$,
that are valid over the whole temperature range $T<T_{\rm c}$ for both
types of pairing state. 

Table \ref{tab:neutrino} summarizes all possible direct Urca processes
with the corresponding relative emissivities. Also included are the
critical density $n_{\rm c}$ and the threshold mass of a non-rotating
neutron star $M_{\rm th}$ above which the respective process is
allowed for two representative EOS's (see
Sect. \ref{sect:eos}). Besides the direct nucleon Urca process ${\rm
n}\rightarrow{\rm p}+{\rm e}^-+\bar\nu$, the most important processes
are $\Sigma^-\rightarrow\Lambda+{\rm e}^-+\bar\nu$,
$\Lambda\rightarrow{\rm p}+{\rm e}^-+\bar\nu$,
$\Sigma^-\rightarrow{\rm n}+{\rm e}^-+\bar\nu$, and
$\Xi^-\rightarrow\Lambda+{\rm e}^-+\bar\nu$. Other processes
are only possible at very high densities just below or above the
maximum central density of a stable hyperon star. As a result, the gap
energy of the $\Lambda$ particle is the most important ingredient
(following those of the nucleons) for simulating the cooling history
of the star, whereas the gap energies of the other hyperons need to be
known only for the most massive stars.  The corresponding modified
hyperon Urca processes obtained by adding a bystander baryon are only
important if no enhanced process is allowed since their emissivities
are considerably smaller (see \cite{Maxwell87a}).

\section{Superfluidity of Hyperons} \label{sect:gap}

The presence of an attractive two-body $\Lambda-\Lambda$ interaction
is implied by doubly-strange hypernuclei, where two $\Lambda$ hyperons
are trapped in a single nucleus (\cite{KeK}). The properties of the
decay products of such hypernuclei indicate that the separation energy
of the two $\Lambda$ hyperons, $B_{\Lambda\Lambda}$, is larger than
twice the separation energy of a single $\Lambda$ from the same core
nucleus, $2B_\Lambda$. The bond energy derived from these experiments
is $\Delta 
B_{\Lambda\Lambda}\!\equiv\!B_{\Lambda\Lambda}-2B_\Lambda\approx4-5\;$
MeV, somewhat less than the corresponding bond energy of nucleons in
nuclei, $B_{NN}\approx6-7\;$MeV. Hence, $\Lambda$ hyperon pairing
similar to that of nucleons is indeed expected in high density matter.

In our simulations presented below we use the $\Lambda$ hyperon \sfs
gaps as estimated recently by Balberg \& Barnea (1998). Their
$\Lambda\Lambda$ potential is based on the $G$-Matrix parameterization
(\cite{Lanskoy97a}), which accurately reproduces the bond energy of 
various doubly strange hypernuclei. In this model, the pairing energy 
is calculated as a function of the $\Lambda$ Fermi momenta and the 
background baryon density, while neglecting relativistic and polarization 
effects.

In general, $\Lambda$ pairing was found to extend from the threshold 
density for $\Lambda$ formation to the density where the 
$\Lambda$ Fermi momenta exceeds $1.3\;$fm$^{-1}$ (typically corresponding 
to a $\Lambda$ fraction of $0.15\!-\!0.2$).  The corresponding gap energies 
were found to be several tenths of a MeV, where the exact 
value depends on both the $\Lambda$ fraction and the background matter
density. These values are actually similar to those found for 
proton \sfs pairing with a neutron matter background (\cite{Elgaroy96c}), 
again indicating the general similarity between hyperon-related and nucleon
related interactions expected at supernuclear densities.

The pairing gaps are also strongly dependent
on the effective mass of the particles on the Fermi surface, since the
pairing interaction competes with the particles' kinetic
energy. Relativistic calculations typically predict that at supernuclear 
densities the the baryon effective masses are considerably lower than the 
bare masses, but the exact value on the Fermi surface is highly model 
dependent. We thus treat the value of the effective mass as a parameter, 
which allows to examine the sensitivity of our results to the uncertainty in
the $\Lambda$ gap energies.

\begin{figure}[tb]
\psfig{figure=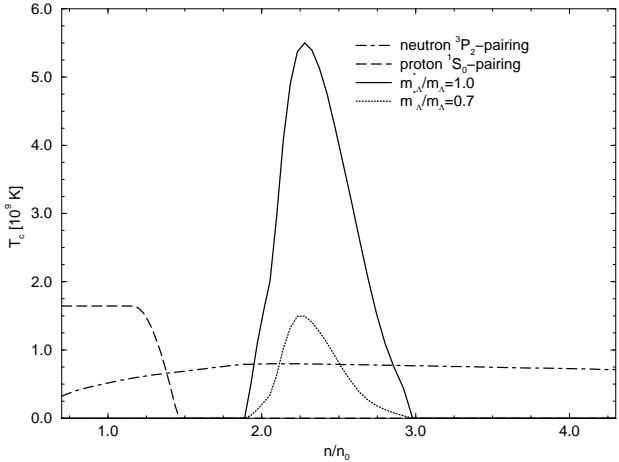,height=\linewidth,angle=-90}
\caption[]{Critical temperatures for the TM1-m1 EOS.
  \label{fig:gap}}
\end{figure}
Figure \ref{fig:gap} shows the critical temperature for $\Lambda$ \sfs
pairing expected for the equilibrium compositions found with the
TM1-m1-model EOS (\cite{Schaffner95a}). The critical
temperature $T_{\rm c}$ is related to the gap energy $\Delta(0)$ at
zero temperature by $T_{\rm c}=a\Delta(0)/k_{\rm B}$, where the
proportional factor $a$ depends on the pairing type (see
\cite{Levenfish94a}). Note the strong dependence of the critical
temperature on the $\Lambda$ effective mass, $m^*_\Lambda$, (unlike
the density range in which \sfs paring exists, which is practically
independent of $m^*_\Lambda$). We show the critical temperatures when
the $\Lambda$ mass on the Fermi surface is taken as equal to the bare
one, $m^*_\Lambda/m_\Lambda=1$, and also when this mass is reduced by
a factor of $m^*_\Lambda/m_\Lambda=0.7$. The latter case is a
reasonable representation for the value of the bulk effective mass as
found in the selfconsistent calculation of the TM1-m1 model. We stress
again that the mass on the Fermi surface may differ considerably from
the bulk effective mass, and the two cases will be used here mainly to
examine the dependence of the thermal history on the $\Lambda$
critical temperature.  Also shown are the critical temperatures for
proton \sfs and neutron \sfp pairing, as calculated by Wambach
et~al. (1991) and Amundsen \& {\O}stgaard (1985), respectively. Note
that neutron \sfp pairing exists over the entire supernuclear density
range in this specific model.

\section{Equations of State} \label{sect:eos}

The EOS is quite unknown at high density. For the present study, we
choose two recent relativistic parameterizations, which have been
proven to be able to describe properties of nuclei and/or nucleon
scattering data.

The first set (TM1-m1) is a relativistic mean-field calculation
(model 1 of \cite{Schaffner95a}). The nucleon
parameters are adjusted to properties of nuclei (set TM1, see
\cite{Sugahara94a}).
The vector potential is nonlinear in density to get a soft behavior at high
density in
accordance with Dirac-Br\"uckner calculations. The hyperon coupling
constant are fixed by SU(6) symmetry and hypernuclear data.

The other set (RHF8 from \cite{Huber97a}) uses
Relativistic-Br\"uckner-Hartree-Fock results up to 2-3 times normal
nuclear density. Hyperons are implemented within the
Relativistic-Hartree-Fock approach.  The EOS calculated within
Relativistic-Hartree-Fock are usually much softer than those derived
from relativistic mean-field models due to the additional degrees of
freedom.  Again, SU(6) symmetry as well as hypernuclear data are used
to fix the hyperon coupling constants.

The hyperons appear at a slightly higher density for RHF8 than in TM1-m1
except for the $\Sigma^-$ which appears around the same density of
$2\rho_0$. The maximum fraction of $\Sigma^-$ is higher in RHF8
than in TM1-m1, while the $\Lambda$ fraction is about a factor of two smaller.
This has an
impact on the onset of the various hyperon cooling processes as seen
in Tab. 1. The critical density for hyperon induced cooling processes
is usually lower for TM1-m1 than for RHF8, except for the process
$\Sigma^-\rightarrow{\rm n}+{\rm e}^-+\bar\nu$, which is
however suppressed by neutron pairing.

\section{Results} \label{sect:results}

\begin{figure}[tb]
\psfig{figure=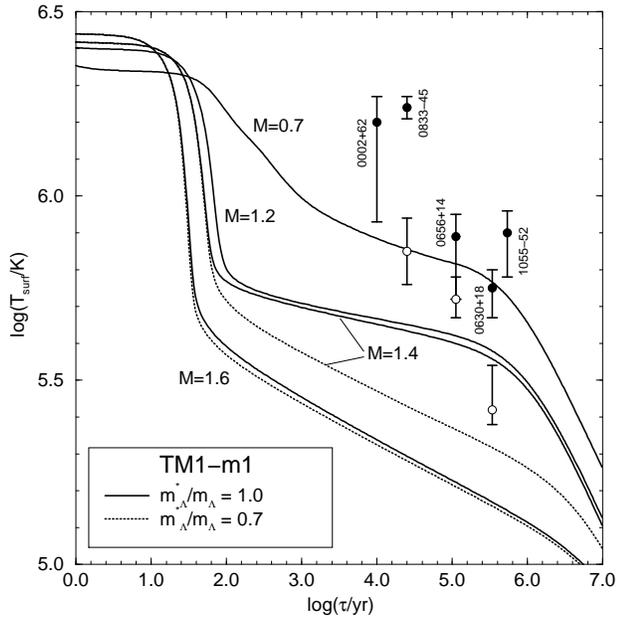,width=\linewidth}
\caption[]{Cooling of neutron and hyperon stars constructed
  for the TM1-m1 EOS. The observational data obtained
  within the magnetic hydrogen atmosphere model are marked by hollow
  circles, whereas the one obtained within the pure iron atmosphere
  model are marked by solid ones. \label{fig:cool1}}
\end{figure}

\begin{figure}[tb]
\psfig{figure=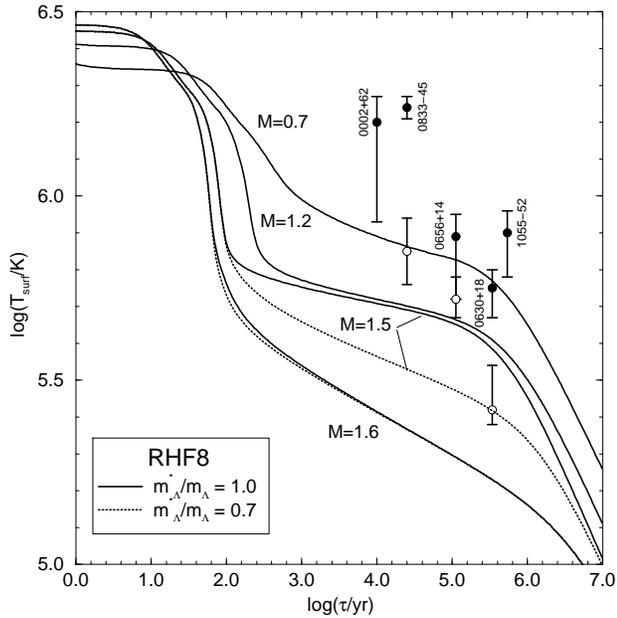,width=\linewidth}
\caption[]{Same as Fig. \ref{fig:cool1} but for the RHF8
  EOS.  \label{fig:cool2}}
\end{figure}

The general relativistic equations of stellar structure and thermal
evolution (see \cite{Thorne77}) were numerically solved via an
implicit finite difference scheme by a Newton-Raphson algorithm (see
\cite{Schaab95a} for more details)\footnote{Tables with detailed 
references to the used ingredients, the used observational data, and
the obtained cooling tracks can be found on the Web:
http://www.physik.uni-muenchen.de\hspace{0pt}/sektion\hspace{0pt}/suessmann\hspace{0pt}/astro\hspace{0pt}/cool\hspace{0pt}/schaab.0897.}. The
surface temperatures deduced from the observed thermal X-ray spectra
depend on the used atmosphere model. We plot the surface temperatures
along with their $2\sigma$ error bars for the two extreme cases of a
pure iron atmosphere (solid circles in Figs. \ref{fig:cool1} and
\ref{fig:cool2}) and of a magnetic hydrogen atmosphere (hollow
circles, see \cite{Schaab98a} for more details). The deduced surface
temperatures can only be understood as upper limits to the evolution
temperature, since magnetospheric emission may irradiate the star's
surface. On the other hand, the magnetic hydrogen atmosphere model is
likely to underestimate the actual surface temperature
(\cite{Potekhin96c}). The two extreme values of the respective star's
surface temperature should therefore give a good estimate of the
maximum range of the true evolution temperature.

Fig. \ref{fig:cool1} shows the cooling behavior of hyperon stars
constructed for the TM1-m1 EOS. The surface temperature
as measured by a distant observer is plotted against the star's age
for different star masses (a larger mass corresponds to a larger central 
density, thus crossing the thresholds of more cooling processes; 
see \cite{Lattimer91}).

\begin{itemize}
\item $M=0.7 M_\odot$: This model is an example for a slow cooling
neutron star. The interior does not contain hyperons, nor is the
proton fraction sufficiently high to allow for the direct nucleon
Urca. Note that the relatively low surface temperature is caused by
the inclusion of the superfluid pair breaking and formation process
(\cite{Voskresenskii87a,Schaab95b}), which is not included in former
works (e.g. \cite{Umeda94,Page95,Schaab95a}).
\item $M=1.2 M_\odot$: The interior consists still only of
nucleons, but the nucleon direct Urca is now possible. This yields a
faster cooling which is slowed down by neutron pairing below the
critical temperature $T_{\rm c}\approx 8\times 10^8$~K.
\item $M=1.4 M_\odot$: At $n\approx 1.8-1.9n_0$ $\Sigma^-$ and
$\Lambda$ hyperons begin to populate the central region of the
star. Besides the direct nucleon Urca the processes
$\Sigma^-\rightarrow\Lambda+{\rm e}^-+\bar\nu$ and
$\Lambda\rightarrow{\rm p}+{\rm e}^-+\bar\nu$ are also possible.  The
two direct hyperon processes are suppressed by $\Lambda$ pairing below
the critical temperature $T_{\rm c}$. At the outer boundary of the
hyperon core $T_{\rm c}$ is rather small ($\approx 10^8$~K for
$m^*_\Lambda/m_\Lambda=1$ and $\approx 10^7$~K for
$m^*_\Lambda/m_\Lambda=0.7$). The hyperon processes are therefore
significantly suppressed only in the innermost part of the star.  The
surface temperature of a middle aged, enhanced cooling neutron star 
($10^2{\rm yrs}\lesssim\tau\lesssim 10^5{\rm yrs}$) is directly related to 
the critical temperature of its superfluids 
(\cite{Lattimer94,Page95,Schaab95a}) and 
depends therefore on the considered gap energy model.  For the $\Lambda$ 
gap model with $m^*_\Lambda/m_\Lambda=1$, the critical temperature of 
$\Lambda$ pairing is larger than for neutron \sfp pairing. In this specific
model, the cooling behavior is therefore determined by the critical
temperature for the latter.
\item $M=1.6 M_\odot$: At densities $n\gtrsim 3n_0$ the scalar 
$\Lambda\Lambda$ interaction becomes repulsive in our model TM1-m1,
and the $\Lambda$'s no longer pair in a \sfs state. The direct
hyperon Urca processes are therefore unsuppressed and the interior of
the star cools very rapidly. This result relies however on the
specific assumptions that we neglected pairing of other hyperons
($\Sigma^{0,\pm}$, $\Xi^{0,-}$) and pairing of $\Lambda$ in a
\sfp state. If such further pairing is allowed the cooling behavior
should be similar to that of a $M=1.4M_\odot$ star.
\end{itemize}

Fig. \ref{fig:cool2} shows the cooling behavior for the RHF8 EOS.
As for TM1-m1 we have four scenarios corresponding to the
four star masses. The cooling behavior of these scenarios are quite
similar for both cases of EOS's, although some aspects
are somewhat model dependent (TM1-m1 stars generally cool more
rapidly).  We emphasize the effect of $\Lambda$ pairing on the thermal
evolution for both EOS's, which is explicit in the
difference between the $M=1.4 M_\odot$ ($M=1.5 M_\odot$ for RHF8) 
thermal sequences and the $M=1.6 M_\odot$ sequences, as the
latter essentially assume no hyperon pairing. This effect is enhanced,
of course, for larger $\Lambda$ pairing energies.

\section{Conclusions and Discussion} \label{sect:conclusion}

In this \emph{Letter} we have studied the effect of hyperon pairing on the 
thermal evolution of hyperon stars. Our general finding is that 
$\Lambda$-hyperon pairing along with nucleon pairing is sufficient 
to suppress fast hyperon-induced cooling processes in all but the most 
massive neutron stars, in a similar fashion as the suppression of 
fast nucleon processes by nucleon pairing. Pairing of other hyperons 
becomes important only for cooling of the most massive stars with 
$M>1.6\;M_\odot$.

As a result of this suppression, the surface temperatures obtained in
cooling simulations of hyperon stars are considerably higher than
those found for the same stars when the $\Lambda$ hyperons are taken
to be in a normal state. In particular, this allows better
compatibility with surface temperatures deduced within the magnetic
hydrogen atmosphere model from observed thermal emission from pulsars,
with which cooling of hyperon stars through unsuppressed rapid cooling
has been previously found to be inconsistent
(\cite{Haensel93,Schaab95a}).

Our quantitative results demonstrate a significant sensitivity of the
surface temperatures to the modeling of the hyperon pairing. We also
call attention to the dependence of the surface temperature on the
details of the high density EOS through the equilibrium
concentrations of the baryon species. In this sense, comparison of
the outcome of theoretical cooling models with soft X-ray and extreme
UV observations provides a powerful tool for investigating the
interiors of pulsars. Future measurement of pulsar thermal emission
may indeed offer valuable indication regarding the physics and
composition of very-high density matter, although much improvement is
still necessary in both theory and observation.

Our main conclusion is that that the existence of hyperons in neutron
stars cannot be excluded by comparing their predicted thermal history
with actual observations from pulsars, provided the hyperons are
allowed to form a superfluid state - similar to that of the
nucleons. Though we have neglected several effects, such as medium
effects on the modified Urca process, possible meson condensation and
accreted envelopes, we believe this main conclusion to be quite
robust.  The study of these effects, as well as pairing of other
hyperon species and other hyperon coupling models will be addressed in
future work.

\begin{acknowledgements}

Ch.~S. gratefully acknowledges the Bavarian State for financial
support. J.~S.B. is supported by the Alexander von Humboldt-Stiftung
and by the U.S. Department of Energy under Contract No.\
DE-AC03-76SF00098. 

\end{acknowledgements}


\end{document}